\documentclass[12pt,thmsa]{article}
\usepackage{amssymb}

\usepackage{sw20aip}

\input{tcilatex}
\begin{document}

\author{Emilio Santos \\
Departamento de F\'{i}sica. Universidad de Cantabria. Santander. Spain}
\title{A hidden variables model for non-relativistic quantum mechanics in terms of
probabilities of particle paths. }
\date{May, 2, 2016 }
\maketitle

\begin{abstract}
Ii is proved that in non-relativistic quantum mechanics (without spin) the
transition probability may be described in terms of particle paths, every
path having a (positive) probability. This leads to a stochastic hidden
variables theory providing an intuitive picture of particle motion. The
change of velocity at every time has a probability that depends on the
potential over a large region, at a difference with the local action of
classical dynamics. Thus the hidden variables theory is non-local like
Bohm's, but not deterministic.
\end{abstract}

\section{Introduction}

In quantum mechanics the state of a physical system is represented by the
wavefunction (more generally the statevector). Usually from the wavefunction
we cannot predict the actual outcomes of the experiments but only the
probabilities of several possible results. Since the early days of quantum
mechanics a controversy has existed about the correct explanation of this
fact. Indeed this was the main subject of the celebrated debate between
Einstein and Bohr, that culminated with the EPR paper\cite{EPR} and
Bohr\'{}s prompt reply\cite{BohrEPR}. Einstein et al. supported the view
that the statistical character of the quantum predictions is due to the fact
that the wavefunction should be associated to an ensemble of states, that
would represent the actual state of the physical system, incompletely known.
In contrast Bohr and his followers supported the completeness of quantum
mechanics, that is the opinion that the wavefunction gives a complete
description of the state. If we accept completeness then we should assume
that the laws of physics are not strictly causal, something that Einstein
disliked. In contrast the incompleteness hypothesis suggests the possibility
of introducing additional parameters in order to define the state of the
physical system in more detail\cite{FOS}. Those parameters have been known
as ``hidden variables''. The mainstream of the scientific community did not
support hidden variables, viewed as useless or even impossible. The latter
belief was reinforced by the von Neumann theorem of 1932 against hidden
variables. However 20 years later David Bohm\cite{Bohm} proposed a specific
hidden variables model, recovering old ideas of de Broglie (the particle is
guided by a wave) and Madelung (the hydrodynamical model of
Schr\"{o}dinger's equation). Bohm\'{}s model proved that hidden variables
are indeed possible, which showed that von Neumann theorem rested upon too
restrictive assumptions. The subject was clarified in two celebrated papers
by John Bell\cite{Bell},\cite{Bell1}. In any case the debate about hidden
variables is not just an old historical fact, but an alive subject as shown
by the effort made to refute empirically at least a class of hidden
variables, namely those local, something achieved only recently\cite{Hensen}%
, \cite{Shalm}, \cite{Giustina}. In this paper I propose a new, nonlocal,
hidden variables theory.

In Bohm's model every particle has an actual path which is completely
determined by a guiding complex field $\psi \left( \mathbf{r},t\right) $
once the initial position is fixed. The field $\psi $ evolves according to
Schr\"{o}dinger equation, that is for a single particle, 
\begin{equation}
i
\rlap{\protect\rule[1.1ex]{.325em}{.1ex}}h%
\frac{\partial \psi }{\partial t}=\hat{H}\psi \equiv \left[ -\frac{%
\rlap{\protect\rule[1.1ex]{.325em}{.1ex}}h%
^{2}}{2m}\nabla ^{2}+V\left( \mathbf{r}\right) \right] \psi ,  \label{2.5}
\end{equation}
$\nabla ^{2}$ being the Laplacian operator and $V\left( \mathbf{r}\right) $
the potential.\textrm{\ }Eq.$\left( \ref{2.5}\right) $ may be separated in
two real fields $R\left( \mathbf{r},t\right) $ and $S\left( \mathbf{r}%
,t\right) $, that is 
\begin{equation}
\psi =R\exp iS/
\rlap{\protect\rule[1.1ex]{.325em}{.1ex}}h%
,  \label{1.1}
\end{equation}
whence eq.$\left( \ref{2.5}\right) $\ becomes a couple of real equations. A
hidden variables model is obtained by introducing Bohm's assumption that the
path, $\mathbf{x}\left( t\right) ,$ of the actual particle is determined by
the ``guiding'' relation 
\begin{equation}
\frac{d\mathbf{x}\left( t\right) }{dt}=\frac{1}{m}\nabla S\left( \mathbf{r,}%
t\right) \mid _{\mathbf{r=x}}.  \label{1.4}
\end{equation}

Several authors have developped Bohm\'{}s model in different directions\cite
{Holland}. Asides from applications (e. g. description of chemical reactions
in terms of molecular paths) some authors consider the model a good picture
of the quantum behaviour. However there are difficulties for that
interpretation, mainly the following:

1) The model is deterministic. This fact may be a good feature for some
people, but it is unable to explain the probabilistic character of quantum
mechanics. Thus one is compelled to introduce Born's rule as an additional
assumption. Indeed there is no reason to use a statistical ensemble when
there is a single particle and the motion is deterministic. Also there is no
clue as to why the real part of the field should be the (square root of) the
probability density in the ensemble.

2) The particle\'{}s law of motion appears as a modification of classical
mechanics by the addition of a kind of force deriving from a ``quantum
potential''. That potential has the strange feature that the force depends
on the gradient of the logaritm of the field $R\left( \mathbf{r},t\right) ,$
whence it may be large even in regions where that field intensity is very
small.

3) In any stationary state, where the wavefunction $\psi $ may be taken as
real and therefore $S=0$, the particle is at rest according to eq.$\left( 
\ref{1.4}\right) $. Nevertheless it is assumed that the position is random
with a probability density determined by the field $R\left( \mathbf{r}%
\right) $. This looks somewhat inconsistent.

In my opinion these difficulties prevent us to consider Bohm\'{}s model as
an appropriate picture for the quantum behaviour. In this paper I propose a
different hidden variables model that eliminates the difficulties, I hope.
It rests on Feynman's path integral formulation of quantum mechanics. In our
formalism the central role is played by the transition probability, rather
than the transition amplitude. It is true that going from the amplitude to
the probability some information is lost, but still some interesing quantum
properties may be calculated, e. g. scattering cross sections. However I
leave the study of applications for future articles.

\section{Transition probability in terms of particle paths}

Feynman\'{}s paths integrals have been very important both from the
fundamental and from the practical points of view\cite{Feynman}. But the
formalism does not provide an intuitive picture of the quantum evolution. It
is the purpose of this paper to show that in nonrelativistic quantum
mechanics of particles without spin (QM\ in the following) it is possible to
get from Feynman's formalism an approach where every particle follows a path
and there is a\textit{\ positive} probability for every possible path.

The fundamental solution of Schr\"{o}dinger eq.$\left( \ref{2.5}\right) $ is
obtained solving it with the initial condition 
\begin{equation}
\psi \left( \mathbf{r,}t_{0}\right) =\delta ^{3}\left( \mathbf{r-r}%
_{0}\right) ,  \label{2.6}
\end{equation}
where $\delta ^{3}\left( \mathbf{r-r}_{0}\right) $ is a three-dimensional
Dirac delta. The solution provides a kernel $A\left( \mathbf{r}_{0}\mathbf{,}%
t_{0}\rightarrow \mathbf{r,}t\right) ,$ such that the solution at time $t$
with a general initial condition $\psi _{0}(\mathbf{x,}t_{0})$ may be
obtained via 
\begin{equation}
\psi \left( \mathbf{r,}t\right) =\int d\mathbf{x}\psi _{0}(\mathbf{x,}%
t_{0})A\left( \mathbf{x,}t_{0}\rightarrow \mathbf{r,}t\right) ,  \label{2.7}
\end{equation}
where the integral is understood to be 3-dimensional. Thus any formalism
that allows getting the kernel (or transition amplitude) $A$ may be taken as
the basis for QM.

Feynman\cite{Feynman} showed that an appropriate kernel (or transition
amplitude), $A_{F}\left( x_{a},t_{a}\rightarrow x_{b},t_{b}\right) $, for
Schr\"{o}dinger eq.$\left( \ref{2.5}\right) $ is the following 
\begin{eqnarray}
A_{F} &=&\lim_{\varepsilon \rightarrow 0}\left( \frac{1}{2\pi i\varepsilon }%
\right) ^{n/2}\int dx_{n-1}...\int dx_{1}\prod_{j=1}^{n-1}\exp \left(
-\gamma x_{j}^{2}\right) ,  \nonumber \\
&&\prod_{j=1}^{n}\exp \left\{ \frac{i}{2\varepsilon }\left(
x_{j}-x_{j-1}\right) ^{2}-\frac{i\varepsilon }{%
\rlap{\protect\rule[1.1ex]{.325em}{.1ex}}h%
}V(\frac{x_{j}+x_{j-1}}{2})\right\} ,  \label{2.9}
\end{eqnarray}
where $\int dx_{j}$ are integrals over the whole real line and $\gamma >0$
is a parameter introduced in order to regularize the integrals. The time
interval $\varepsilon \equiv t_{j}-t_{j-1},$ is independent of $j$ and $%
x_{0}\equiv x_{a},x_{n}\equiv x_{b}.$ The limit $\varepsilon \rightarrow 0$
should be understood with $n\rightarrow \infty $ fulfilling 
\begin{equation}
n\varepsilon =t_{b}-t_{a}.  \label{nepsilon}
\end{equation}
From now I will work in one dimension and use units such that $
\rlap{\protect\rule[1.1ex]{.325em}{.1ex}}h%
=m=1$.

Actually other kernels different from $A_{F}$ eq.$\left( \ref{2.9}\right) $
may be used. They would differ from Feynman's by terms of order $O\left(
\varepsilon \right) .$ In this paper I will start using another one
substituting $\left[ V(x_{j-1})+V(x_{j})\right] /2$ for $\left[ V\left(
x_{j-1}+x_{j-1}\right) /2\right] $ or, what is equivalent to order $%
\varepsilon ,$ writing the amplitude in the form 
\begin{eqnarray}
A(a &\rightarrow &b)=\lim_{\varepsilon \rightarrow 0}\left( \frac{1}{2\pi
i\varepsilon }\right) ^{n/2}\int dx_{n-1}...\int dx_{1}\prod_{j=1}^{n-1}\exp
\left( -\gamma x_{j}^{2}\right)  \nonumber \\
&&\times \prod_{j=1}^{n}\exp \left\{ \frac{i}{2\varepsilon }\left(
x_{j}-x_{j-1}\right) ^{2}-i\varepsilon V(x_{j})\right\} .  \label{2.14}
\end{eqnarray}
I omit the proof that eq.$\left( \ref{2.14}\right) $ leads to the correct
probability amplitude, solution of Schr\"{o}dinger eq.$\left( \ref{2.5}%
\right) ,$ which is almost identical to the one for eq.$\left( \ref{2.9}%
\right) $\cite{Feynman}.

In QM the transition probability is the square modulus of the transition
amplitude whence we get, taking eq.$\left( \ref{2.14}\right) $ into account, 
\begin{eqnarray}
P(a &\rightarrow &b)\equiv P(z_{a},t_{a}\rightarrow z_{b},t_{b})=\left|
A(a\rightarrow b)\right| ^{2}  \nonumber \\
&=&\lim_{\varepsilon \rightarrow 0}\left( \frac{1}{2\pi i\varepsilon }%
\right) ^{n}\prod_{j=1}^{n-1}\int dx_{j}\int dy_{j}\prod_{j=1}^{n-1}\exp
\left( -\gamma x_{j}^{2}-\gamma y_{j}^{2}\right)  \nonumber \\
&&\times \prod_{j=1}^{n}\exp \left\{ \frac{i}{2\varepsilon }\left[ \left(
y_{j}-y_{j-1}\right) ^{2}-\left( x_{j}-x_{j-1}\right) ^{2}\right] \right\} 
\nonumber \\
&&\times \prod_{j=1}^{n-1}\exp \left\{ i\varepsilon \left[
V(x_{j})-V(y_{j})\right] \right\} .  \label{3.2}
\end{eqnarray}
and we have taken into account that $x_{0}=y_{0}=z_{b},x_{n}=y_{n}=z_{b}.$
The quantity $P(a\rightarrow b)$ has dimensions of probability per square
volume and it should be interpreted as a relative probability.

Now I make a change of variables, that is 
\begin{equation}
z_{j}=\frac{1}{2}\left( x_{j}+y_{j}\right) ,u_{j}=x_{j}-y_{j},0\leq j\leq n,
\label{newvar}
\end{equation}
whence eq.$\left( \ref{3.1}\right) $ becomes, reordering the integrals and
the exponentials, 
\begin{eqnarray}
P(a &\rightarrow &b)=\lim_{\varepsilon \rightarrow 0}\left( \frac{1}{2\pi
\varepsilon }\right) ^{n}\prod_{j=1}^{n-1}\int dz_{j}\int du_{j}\exp \left[
-\gamma z_{j}^{2}-\gamma u_{j}^{2}\right]  \nonumber \\
&&\times \exp \left\{ -iu_{j}s_{j}+i\varepsilon \left[ V(z_{j}-\frac{1}{2}%
u_{j})-V(z_{j}+\frac{1}{2}u_{j})\right] \right\} ,\smallskip  \label{3.3}
\end{eqnarray}
where $z_{0}=z_{a}$, $z_{n}=z_{b}$ and $u_{0}=u_{n}=0$ and 
\begin{equation}
s_{j}\equiv \frac{z_{j+1}-2z_{j}+z_{j-1}}{\varepsilon },j=1,2,...n-1.
\label{7.14b}
\end{equation}

Eq.$\left( \ref{3.3}\right) $ may be written\textrm{\ } 
\begin{equation}
P(a\rightarrow b)=\frac{1}{2\pi \left( t_{b}-t_{a}\right) }\lim_{\varepsilon
\rightarrow 0}\int dz_{1}...\int dz_{n-1}n\prod_{j=1}^{n-1}Q_{j},\smallskip
\label{4.20}
\end{equation}
where 
\begin{eqnarray}
Q_{j} &=&\left( 2\pi \varepsilon \right) ^{-1}\exp \left[ -\gamma
z_{j}^{2}\right] \int du_{j}\exp \left[ -\gamma u_{j}^{2}-iu_{j}\cdot
s_{j}\right]  \nonumber \\
&&\times \exp \left\{ i\varepsilon \left[ V(z_{j}-\frac{1}{2}u_{j})-V(z_{j}+%
\frac{1}{2}u_{j})\right] \right\} .  \label{6.1}
\end{eqnarray}

The convergence factors $\exp \left[ -\gamma z_{j}^{2}-\gamma
u_{j}^{2}\right] $ derive from the choice of regularization made in eq.$%
\left( \ref{3.3}\right) $. Indeed the choice has the virtue of simplicity,
but it is not appropriate for our purposes. We need a more slow
regularization factor and I will replace eq.$\left( \ref{6.1}\right) $ by
the following one 
\begin{eqnarray}
Q_{j} &=&\left( 2\pi \varepsilon \right) ^{-1}\exp \left[ -\gamma \left|
z_{j}\right| \right] \int du_{j}\exp \left[ -\gamma \left| u_{j}\right|
-iu_{j}\cdot s_{j}\right]  \nonumber \\
&&\times \exp \left\{ i\varepsilon \left[ V(z_{j}-\frac{1}{2}u_{j})-V(z_{j}+%
\frac{1}{2}u_{j})\right] \right\} .  \label{6}
\end{eqnarray}

Thus eq.$\left( \ref{4.20}\right) $ gives the quantum transition
probability, $P(z_{a},t_{a}\rightarrow z_{b},t_{b}),$ in the form of a path
integral, every path corresponding to one possible motion of the particle
starting in ($z_{a},t_{a})$ and finishing in ($z_{b},t_{b})$. For any finite
value of $n$ (and $\varepsilon =\left( t_{b}-t_{a}\right) /n)$ every path is
defined by $n+1$ spacetime points $\left\{ z_{j},t_{j}\right\} ,$ with the
assumption that the motion in every time interval $\left\{ j\varepsilon
,(j+1)\varepsilon \right\} $ is uniform. The ``weight'' 
\begin{equation}
W_{n}\left( \left\{ z_{j}\right\} \right) \equiv n\prod_{j=1}^{n-1}Q_{j},
\label{6.2}
\end{equation}
plays the role of the (relative) probability of the path \textit{provided
that }$W_{n}\geq 0$\textit{. } It is trivial to show that the quantity $%
W_{n} $ is real. In fact any $Q_{j}$, eq.$\left( \ref{6.1}\right) ,$ is real
because the imaginary part of the integrand does not contribute, it having
the wrong symmetry. However there may be paths whose ``weight'' $%
W_{n}\left\{ z_{j}\right\} $ is negative for some potentials $V(x)$. In the
following I show that for a restricted class of potentials weight of every
path is positive (or zero) provided that the path is defined with a large
enough $n.$ Nevertheless the restricted class of pontentials cover all those
physically sensible, as shown in the following.

\section{Positivity of the path weights}

The proof that the weights eq.$\left( \ref{6.2}\right) $ are positive rests
upon the following

\begin{theorem}
If the Fourier transform of the potential $V(x)$ has compact support and its
integral over its domain is bounded, there is a positive number $\lambda >0$
such that the weights $W_{n},$ eq.$\left( \ref{6.2}\right) ,$ are
nonnegative for all paths $\left\{ z_{j}\right\} $ involving a number, $n+1$%
, of points that fulfils $n>\left( t_{b}-t_{a}\right) /\lambda .$
\end{theorem}

The assumptions of the theorem about the potential may be written 
\begin{equation}
\tilde{V}\left( q\right) =0\text{ for }\left| q\right|
>R,\int_{-R}^{R}\left| \tilde{V}\left( q\right) \right| dq\leq K,
\label{3.5}
\end{equation}
where $K$ and $R$ are positive parameters and $\tilde{V}\left( q\right) $ is
the Fourier transform of the potential, that is 
\[
\tilde{V}\left( q\right) \equiv \int V\left( x\right) \exp \left( ixq\right)
dx\mathbf{.} 
\]
Of course in a nonrelativistic theory the zero of energies may be fixed
arbitrarily so that $V=const$ is physically equivalent to $V=0$. Thus the
latter eq.$\left( \ref{3.5}\right) $ should be understood modulo an
appropriate redefinition of the zero of the potential.

Before going to the proof I will comment on the rationale for the hypotheses
of the theorem. We are interested in the positivity of the weights $W_{n}$
only in the limit $n\rightarrow \infty $ (or $\varepsilon \rightarrow 0),$
and this is guaranteed if $W_{n}$ is positive for any $n$ large enough. On
the other hand the constraints eq.$\left( \ref{3.5}\right) $ guarantee that
the classical force is finite at any point, a rather obvious physical
requirement. In fact the force on the particle at the point $x$ is given by 
\begin{eqnarray*}
F\left( x\right) &=&-dV\left( x\right) /dx=-\frac{d}{dx}\left[ \frac{1}{2\pi 
}\int \tilde{V}\left( q\right) \exp \left( -ixq\right) dq\right] \\
&=&-\frac{i}{2\pi }\int q\tilde{V}\left( q\right) \exp \left( -ixq\right) dq%
\mathbf{,}
\end{eqnarray*}
where we have used the inverse Fourier transform of the latter eq.$\left( 
\ref{3.5}\right) $. Hence the force is bounded, that is 
\[
\left| F\left( x\right) \right| \leq \frac{1}{2\pi }\int_{-R}^{R}\left| q%
\tilde{V}\left( q\right) \right| dq\leq \frac{R}{2\pi }\int_{-R}^{R}\left| 
\tilde{V}\left( q\right) \right| dq\leq \frac{RK}{2\pi }, 
\]
where the former eq.$\left( \ref{3.5}\right) $ has been taken into account.
Thus all physically plausible potentials defined in a finite region of space
are allowed by the constraints eq.$\left( \ref{3.5}\right) $ of our theorem
if $K$ and $R$ are large enough. This includes the constant potential and
the truncated harmonic oscillator 
\[
V(x)=\frac{1}{2}kx^{2}\text{ if }\left| x\right| \leq L,0\text{ otherwise,} 
\]
with $k$ not too large, that is $k<K/(2RL^{2}).$

\textbf{Proof of the theorem}

I start with a change leading to a description of the transition
probability, different but equivalent to eq.$\left( \ref{3.3}\right) $ in
the limit $n\rightarrow \infty .$ I will write

\begin{eqnarray}
P &=&\lim_{n\rightarrow \infty }\left( \frac{1}{2\pi \varepsilon }\right)
^{n}\int dz_{n-1}...\int dz_{1}\int du_{n-1}...\int du_{1}  \nonumber \\
&&\times \prod_{j=1}^{n-1}\exp \left[ -\gamma \left| z_{j}\right| -\gamma
\left| u_{j}\right| \right] \exp \left[ -\frac{i}{\varepsilon }u_{j}\left(
z_{j-1}\mathbf{-}2z_{j}\mathbf{+}z_{j+1}\right) \right]  \nonumber \\
&&\times \prod_{j=1}^{n-1}\left\{ 1+i\varepsilon \left[ V(z_{j}-\frac{1}{2}%
z_{j})-V(z_{j}+\frac{1}{2}z_{j})\right] \right\} .  \label{4.1}
\end{eqnarray}
The proof of equivalence follows from expanding the exponentials in the
latter product of eq.$\left( \ref{3.3}\right) $ in powers of the small
parameter $\varepsilon ,$ that is 
\begin{equation}
\exp \left( i\varepsilon B_{j}\right) =1+i\varepsilon B_{j}-\frac{%
\varepsilon ^{2}}{2}B_{j}^{2}-\frac{i\varepsilon ^{3}}{6}B_{j}^{3}+...,
\label{4.11}
\end{equation}
where for short I have labelled 
\[
B_{j}\equiv V(r_{j}-\frac{1}{2}u_{j})-V(r_{j}+\frac{1}{2}u_{j}). 
\]
The relevant result is that only the term of order $\varepsilon $ in eq.$%
\left( \ref{4.11}\right) $ contributes to eq.$\left( \ref{4.1}\right) $ in
the limit $\varepsilon \rightarrow 0.$ In fact we have 
\begin{equation}
\prod_{j}\left( 1+i\varepsilon B_{j}\right) =1+i\varepsilon
\sum_{j}B_{j}-\varepsilon ^{2}\sum_{l}\sum_{j>l}B_{l}B_{j}+...  \label{4.12}
\end{equation}
As the sum $\sum_{j}B_{j}$ consists of $n-1$ terms, the quantity $%
\varepsilon \sum_{j}B_{j}$ has a finite limit when $\varepsilon \rightarrow
0 $ (remember that we assume $n\varepsilon =t_{b}-t_{a},$ finite in the
limit). Similarly there are $n(n-1)/2$ terms in the double sum $%
\sum_{l}\sum_{j>l}B_{l}B_{j}$ so that its product times $\varepsilon ^{2}$
has also a finite limit. The same happens for every term in the right side
of eq.$\left( \ref{4.12}\right) .$ In sharp contrast the terms containing $%
B_{j}^{s}$ in eq.$\left( \ref{4.11}\right) $ with $s>1$ have extra factors $%
\varepsilon $ whence they do not contribute in the limit $\varepsilon
\rightarrow 0.$ For instance there will be $n\left( n-1\right) $ terms of
order $\varepsilon ^{3}$ not included in the sum eq.$\left( \ref{4.12}%
\right) ,$ namely those of the form 
\[
(i\varepsilon B_{j})\left( -\frac{\varepsilon ^{2}}{2}B_{k}^{2}\right) =-%
\frac{i\varepsilon ^{3}}{2}B_{j}B_{k}^{2}. 
\]
The sum of these terms contributes a quantity of order $n^{2}\varepsilon
^{3} $ that will go to zero in the limit $\varepsilon \rightarrow 0.$ A
similar argument is valid for $s>3$. This completes the proof that eq.$%
\left( \ref{4.1}\right) $ is equivalent to eq.$\left( \ref{4.20}\right) $.

A sufficient condition for the nonnegativity of $W_{n},$eq.$\left( \ref{6.2}%
\right) $ is that $Q_{j}\geq 0$ for any $j$, with $Q_{j}$ now redefined as 
\begin{eqnarray}
Q_{j} &=&\left( 2\pi \varepsilon \right) ^{-1}\exp \left[ -\gamma \left|
z_{j}\right| \right] \int du_{j}\exp \left[ -\gamma \left| u_{j}\right|
-iu_{j}s_{j}\right]  \nonumber \\
&&\times \left\{ 1+i\varepsilon \left[ V(z_{j}-\frac{1}{2}u_{j})-V(z_{j}+%
\frac{1}{2}u_{j})\right] \right\} ,  \label{7}
\end{eqnarray}
where $s_{j}$ was defined in eq.$\left( \ref{7.14b}\right) .$ The potential
may be written in terms of its Fourier transform, which leads to 
\begin{eqnarray*}
V(z_{j}-\frac{1}{2}u_{j})-V(z_{j}+\frac{1}{2}u_{j}) &=&\frac{1}{2\pi }%
\int_{-R}^{R}dq_{j}\left| \tilde{V}\left( q_{j}\right) \right| \exp \left(
-iz_{j}q_{j}\right) \\
&&\times \left[ \exp \left( iu_{j}q_{j}\right) -\exp \left(
-iu_{j}q_{j}\right) \right] ,
\end{eqnarray*}
where I have taken into account the constraint eq.$\left( \ref{3.5}\right) .$
If this is inserted in eq.$\left( \ref{7}\right) $ the $u_{j}$ integrals are
trivial and we get 
\begin{eqnarray}
Q_{j} &=&\left( 2\pi \varepsilon \right) ^{-1}\exp \left[ -\gamma \left|
z_{j}\right| \right] \frac{2\gamma }{s_{j}^{2}+\gamma ^{2}}\left(
1-\varepsilon M_{j}\right) ,  \nonumber \\
M_{j} &=&\pi ^{-1}\int_{0}^{R}dq_{j}\func{Im}\left[ \tilde{V}\left(
q_{j}\right) \exp \left( -iz_{j}q_{j}\right) \right]  \nonumber \\
&&\times \left[ \frac{s_{j}^{2}+\gamma ^{2}}{(s_{j}-q_{j})^{2}+\gamma ^{2}}-%
\frac{s_{j}^{2}+\gamma ^{2}}{(s_{j}+q_{j})^{2}+\gamma ^{2}}\right] ,
\label{8}
\end{eqnarray}
where I have taken into account that $\tilde{V}\left( -q_{j}\right) =\tilde{V%
}^{*}\left( q_{j}\right) $ so that the real part of $\tilde{V}\left(
q_{j}\right) \exp \left( -iz_{j}q_{j}\right) $ does not contribute, and the
imaginary part is odd with respect to $q_{j}.$

Now $Q_{j}$ would be nonnegative if $M_{j}\leq 1/\varepsilon ,$ and this
will happen for any $\varepsilon \leq \lambda $ fulfilling $1/\lambda \geq
M_{j}\left( z_{j},s_{j}\right) $ for all $j$. That is the parameter $\lambda 
$ proposed in the theorem exists if $M_{j}$ is bounded from above. In fact
there is a bound that may be calculated as follows. Taking into account that 
$\gamma <<R$ and $\left| q_{j}\right| \leq R,$we have 
\[
\left| \frac{s_{j}^{2}+\gamma ^{2}}{(s_{j}-q_{j})^{2}+\gamma ^{2}}-\frac{%
s_{j}^{2}+\gamma ^{2}}{(s_{j}+q_{j})^{2}+\gamma ^{2}}\right| \lesssim \frac{%
R^{2}}{\gamma ^{2}},
\]
the maximum corresponding to $\left| s_{j}\right| =\left| q_{j}\right| =R.$
This leads to 
\begin{equation}
M_{j}\leq \left| M_{j}\right| \leq \frac{1}{2\pi }\int_{-R}^{R}dq_{j}\left| 
\tilde{V}\left( q_{j}\right) \right| \frac{R^{2}}{\gamma ^{2}}\leq \frac{%
R^{2}K}{2\pi \gamma ^{2}},  \label{M}
\end{equation}
where the constraints eqs.$\left( \ref{3.5}\right) $ have been taken into
account. As a consequence there is a parameter $\lambda =$ $2\pi \gamma
^{2}/\left( R^{2}K\right) >0$ as stated, that \textit{completes the proof of
the theorem.}

As a consequence of the theorem the quantity $W_{n}\left( \left\{
z_{j}\right\} \right) $ eq.$\left( \ref{6.2}\right) $ is nonnegative in the
limit $n\rightarrow \infty $ for all values of the (real positive) parameter 
$\gamma $. Thus $W_{n}\left( \left\{ z_{j}\right\} \right) $ might be
interpreted as the probability of a path defined by the spacetime points $%
(z_{j},t_{j})$. When $\gamma $ approaches $0$ the quantity $P\left(
a\rightarrow b\right) ,$ eq.$\left( \ref{4.20}\right) ,$ approaches the
quantum transition probability, that is the square of the quantum transition
amplitude defined in eq.$\left( \ref{2.7}\right) $. Thus the quantum
transition probability may be obtained in the form of a functional integral
over particle paths, that may be described as follows 
\begin{equation}
P(a\rightarrow b)=\frac{1}{2\pi \left( t_{b}-t_{a}\right) }\lim_{\gamma
\rightarrow 0}\lim_{\varepsilon \rightarrow 0}n\int dz_{1}...\int
dz_{n-1}\prod_{j=1}^{n-1}Q_{j},  \label{P}
\end{equation}
where $Q_{j}$ was given in eq.$\left( \ref{7}\right) .$ It is necessary that
the limit $\gamma \rightarrow 0$ is taken after the limit $\varepsilon
\rightarrow 0$ in order that the $z_{j}$ integrals are well defined.

The calculations using eq.$\left( \ref{P}\right) $ are rather involved even
for simple potentials. For instance for the free particle, that is $V(x)=0$,
the integral in $z_{1}$ involves the two functions $Q_{1}$ and $Q_{2}$
because $z_{1}$ enters in both, that is 
\begin{eqnarray}
P_{01} &\equiv &\int_{-\infty }^{\infty }Q_{1}Q_{2}dz_{1}=2\left( 2\pi
\varepsilon \right) ^{-2}\int_{0}^{\infty }\exp \left( -\gamma z_{1}\right) 
\nonumber \\
&&\times \frac{4\gamma ^{2}dz_{1}}{\left\{ \left[ \left(
z_{2}-2z_{1}+z_{a}\right) /\varepsilon \right] ^{2}+\gamma ^{2}\right\}
\left\{ \left[ \left( z_{3}-2z_{2}+z_{1}\right) /\varepsilon \right]
^{2}+\gamma ^{2}\right\} }.  \label{Q}
\end{eqnarray}
After this integral is performed we could make the integral in $z_{2}$ and
so on. The calculation is straightforward but lengthy and it will not be
continued here. However the physics of eq.$\left( \ref{Q}\right) $ is clear,
when $\gamma $ is small the function under the integral is picked at 
\[
z_{2}-2z_{1}+r_{a}\simeq z_{3}-2z_{2}+z_{1}\simeq 0\Rightarrow
z_{3}-z_{2}\simeq z_{2}-z_{1}\simeq z_{1}-r_{a}. 
\]
That is the velocity changes but slightly whence the motion becomes close to
classical. Therefore in the formalism the quantum free particle follows the
classical path with probability $1$ in the limit $\gamma \rightarrow 0.$ For
other potentials, $V\left( x\right) \neq const,$ there is some probability
that the velocity changes and the probability depends, via the Fourier
transform of the potential, on some region around the instantaneous position
of the particle. That is \textit{the potential produces an effect which is
stochastic and nonlocal.}

From the mathematical point of view $W_{n}\left( \left\{ z_{j}\right\}
\right) $ is a stochastic process with discrete time for finite $n$, that is
a stochastic chain and becomes continuous in the limit $n\rightarrow \infty $
(with $n\varepsilon =t_{b}-t_{a}).$ The chain (or the continuous) stochastic
process is not Markovian, as shown in eq.$\left( \ref{Q}\right) $ where the
transition probability from $z_{0}$ to $z_{1}$ depends not only on these two
values but also in $z_{2}$ and $z_{3}$. This contrasts with the path
integral formulation of the quantum amplitude, eq.$\left( \ref{2.9}\right) ,$
where every term in the product depends only on the two positions involved, $%
x_{j}$ and $x_{j-1}.$Therefore the Chapman-Kolmogorov equation is not
fulfilled, that is the equality 
\begin{equation}
\int dr_{c}P(r_{a},t_{a}\rightarrow r_{c},t_{c})P(r_{c},t_{c}\rightarrow
r_{b},t_{b})=P(r_{a},t_{a}\rightarrow r_{b},t_{b})  \label{CK}
\end{equation}
does not hold true in general. This is the reason why it is not possible to
get the quantum transition probability as the solution of a Fokker-Planck
equation. In contrast the quantum amplitudes do fulfil an equation similar
to eq.$\left( \ref{CK}\right) ,$ although involving complex amplitudes
rather than real positive probabilities, that is 
\[
\int dr_{c}A(r_{a},t_{a}\rightarrow r_{c},t_{c})A(r_{c},t_{c}\rightarrow
r_{b},t_{b})=A(r_{a},t_{a}\rightarrow r_{b},t_{b}), 
\]
This is consisten with the amplitude being governed by an equation
resembling a diffusion (Fokker-Planck) equation although in complex space,
that is Schr\"{o}dinger eq.$\left( \ref{2.5}\right) .$

\section{Conclusions}

The formalism developped in this paper provides a description of
(non-relativistic) quantum motion more detailed than the standard quantum
description, but in agreement with QM when appropriate averages are made. In
particular, if we add the probabilities of different paths in order to get
the total transition probability, see eq.$\left( \ref{6}\right) .$ Thus the
formalism is a hidden variables theory (HVT) of quantum mechanics. The
theory is stochastic in the sense that it does not provide a deterministic
law of motion (as for instance Bohm's HVT does) but the probability of the
different possible paths of the quantum particle.

A path may be defined by the positions $\left\{ r_{a}\equiv
r_{0},...r_{j},...r_{b}\equiv r_{n}\right\} $ at times $t_{a,}...t_{j}\equiv
t_{a}+j\varepsilon ,...t_{b}$ or, what is equivalent, the end positions plus
the velocity changes $\left\{ s_{j}\right\} $ at times $t_{j}.$ Eventually
we should consider the limit $n\rightarrow \infty $ with $n\varepsilon
=t_{b}-t_{a}.$ These probabilities depend on the potential $V\left( r\right) 
$ along the path, therefore we may say that the potential governs the motion
of the particle, as is shown by eqs.$\left( \ref{11}\right) $ and $\left( 
\ref{14}\right) .$ However the action of the potential is less direct than
in the classical case, in particular the action depends, via the Fourier
transform of the potential, on a whole region around the particle. Thus the
hidden variables model is nonlocal.

I emphasize again that we remain at the level of non-relativistic quantum
mechanics. I do not claim that a similar interpretation may be extended to
relativistic quantum field theory, e. g. photons, electrons when spin plays
a role, or even atoms or molecules when (Bose or Fermi) statistics is
relevant. The existence of pictures for physical theories is considered
irrelevant, even useless, for many people. But for some scientists
``pictures of the reality''\cite{EPR} are an essential part of physics. Also
the HVT may be useful for some calculations where we want to emphasize the
particle aspect of quantum systems, like molecular dynamics.


\begin{thebibliography}{99}
\bibitem{EPR}  A. Einstein, B. Podolsky and N. Rosen, Can quantum-mechanical
description of physical reality be considered complete?. \emph{Phys. Rev. }%
\textbf{47}, 777-780 (1935).

\bibitem{BohrEPR}  N. Bohr, Can quantum-mechanical description of physical
reality be considered complete?. \emph{Phys. Rev. }\textbf{48}, 696 (1935).

\bibitem{Bohm}  D. Bohm, A suggested interpretation of the quantum theory in
terms of ``hidden'' variables, I and II, \textit{Phys. Rev.} \textbf{85},
166-193 (1952).

\bibitem{Bell}  J. S. Bell, On the problem of hidden variables in quantum
mechanics. \emph{Reviews of Modern Physics}, \emph{38}, 447-52 (1966).

\bibitem{Bell1}  J. S. Bell, On the Einstein, Podolsly and Rosen paradox. 
\emph{Physics}, \emph{1}, 195-200 (1964).

\bibitem{FOS}  E. Santos, Towards a realistic interpretation of quantum
mechanics providing a model of the physical world. \textit{Foundations of
Science, DOI 10.1007/s10699-014-9366-y (2015).}

\bibitem{Hensen}  B. Hensen et al., Experimental loophole-free violation of
a Bell inequality using entangled electron spins separated by 1.3 km.
doi.org/10.1038/\textit{nature} 15759 (2015); arXiv: 1508.05949.

\bibitem{Shalm}  L. K. Shalm et al., A strong loophole-free test of local
realism. \textit{Phys. Rev. Lett}. \textbf{115}, 250402 (2015); arXiv:
1511.03189.

\bibitem{Giustina}  M. Giustina et al., A significant loophole-free test of
Bell's theorem with entangled photons. \textit{Phys. Rev. Lett.} \textbf{115}%
, 250401 (2015); arXiv: 1511.03190.

\bibitem{Holland}  P. R. Holland, \textit{The Quantum Theory of Motion}.
Cambridge U. P., Cambridge, U. K., 1993.

\bibitem{Feynman}  R. P. Feynman and A. R. Hibbs, \textit{Quantum Mechanics
and Path Integrals}. Mc.Graw-Hill, New York, 1965.
\end{thebibliography}
\end{document}